\documentclass[12pt]{article}
\usepackage{amssymb,amsmath}
\usepackage{graphicx}
\def\hybrid{\topmargin -20pt  \oddsidemargin 0pt
       \headheight 0pt   \headsep 0pt
       \textwidth 6.25in 
       \textheight 9.5in 
       \marginparwidth .875in
       \parskip 5pt plus 1pt   \jot = 1.5ex}

\hybrid

\def\mC{\mathbb{C}}
\def\mF{\mathbb{F}}
\def\mP{\mathbb{P}}
\def\mR{\mathbb{R}}
\def\mZ{\mathbb{Z}}
\def\zf#1#2{z^{#1}_{{\rm fixed},#2}}

\numberwithin{equation}{section}
\begin{document}

\thispagestyle{empty}
\rightline{MPP-2005-169}
\rightline{DISTA-UPO-05}
\rightline{hep-th/0512287}
\vspace{1truecm}
\vspace{1truecm}
\centerline{\bf \large Moduli Stabilization in Toroidal Type IIB Orientifolds}
\vspace{1truecm}
\centerline{S. Reffert$^{a,}$\footnote{sreffert@theorie.physik.uni-muenchen.de},
E. Scheidegger$^{b,}$\footnote{esche@mfn.unipmn.it}
}

\vspace{.7truecm}

{\em \centerline{$^a$Max--Planck--Institut f\"ur Physik,
F\"ohringer Ring 6, 80805 M\"unchen, Germany}}

\vspace{.4truecm}

\centerline{\em{$^b${Universita del Piemonte Orientale "A. Avogadro",}}}
\centerline{\em{Dip. Scienze e Tecnologie Avanzate,}}
\centerline{\em{Via Bellini, 25/g, I-15100 Alessandria, Italy}}

\vspace{.7truecm}
\vspace{.4truecm}



\begin{abstract}
We discuss the first step in the moduli stabilization program a la KKLT for a general class of resolved toroidal type IIB orientifolds. In particular, we discuss their geometry, the topology of the divisors relevant for the D3--brane instantons which can contribute to the superpotential, and some non--trivial aspects of the orientifold action. \end{abstract}


\def\mC{\mathbb{C}}
\def\mF{\mathbb{F}}
\def\mP{\mathbb{P}}
\def\mR{\mathbb{R}}
\def\mZ{\mathbb{Z}}
\def\zf#1#2{z^{#1}_{{\rm fixed},#2}}
\def\mod{\mathrm{mod}}

\section{Introduction}
\label{sec:intro}

Most string theory compactifications come with a large number of moduli. Corresponding to massless fields in the low energy effective action, they give rise to a fifth force of about gravitational strength and are consequently in conflict with experiment. Therefore, a lot of effort is being dedicated to the subject of moduli stabilization. 

Compactifications which should ultimately yield the MSSM are required to have $N=1$ supersymmetry in four dimensions. One way to achieve this is to start with type II string theory compactified on a compact Calabi--Yau threefold $X$ yielding $N=2$ supersymmetry. Having non--abelian gauge groups requires the addition of D--branes wrapping supersymmetric cycles in $X$. Their presence induces an RR flux which has to be cancelled by adding appropriate orientifold planes. Without D--branes and orientifold planes, the relevant moduli are the dilaton, and the deformations of the complex structure and the K\"ahler form of $X$ while the moduli of the RR fields decouple. After adding the orientifold planes, the geometric and the RR moduli get mixed non--trivially. 

In the KKLT-proposal \cite{Kachru:2003aw}, starting from type IIB theory, the dilaton and the complex structure moduli are stabilized via background three--form fluxes, while the K\"ahler moduli are fixed by non--perturbative effects. In this way, a stable, supersymmetric AdS vacuum can be obtained. In a second step, this vacuum is supposed to be lifted up to a metastable deSitter vacuum by introducing anti--D3--branes, which give rise to a positive contribution to the potential, producing the uplift to a small value of the cosmological constant.

We will be concerned with the first step only, i.e. we study the issue of the stabilization of all moduli. We will present a program for testing this proposal in a number of examples and point out issues which deserve to be understood more deeply in order to fully realize it. This is a continuation of~\cite{Lust:2005dy} and a preview of some of the results of~\cite{toappear}.

The stabilization of the complex structure moduli and the dilaton via three-form fluxes has been extensively discussed in the literature \cite{Giddings:2001yu} (for a review, see \cite{Grana:2005jc} and references therein), while the general conditions for the generation of a non--perturbative superpotential depending on the K\"ahler moduli have recently received a lot of attention, see~\cite{hep-th/0501081,hep-th/0503072,hep-th/0503125,hep-th/0503138,hep-th/0504041,Berglund:2005dm,Bergshoeff:2005yp,hep-th/0507091,Lust:2005cu}.
On the other hand, some concrete realizations of step one of KKLT, i.e. models with all moduli fixed have been put forward~\cite{Denef:2004dm}, \cite{Denef:2005mm}, \cite{Aspinwall:2005ad}.

The closed string moduli space consists of $h^{1,1}(X)$ K\"ahler moduli, which parametrize the size of $X$, $h^{2,1}(X)$ complex structure moduli, parametrizing the shape of $X$, plus the complex dilaton $S=i\,C_0+e^{-\phi}$, which consists of the RR 0--form $C_0$ and the dilaton $\phi$. The three-form flux that is turned on is of the form $G_3=F_3+iS\,H_3$, with $F_3=dC_2$ is the field strength of the RR 2--form $C_2$ and $H_3=dB_2$ is the field strength of the B--field. Such a 3--form flux gives rise to a superpotential~\cite{Gukov:1999ya}, \cite{Taylor:1999ii} of the form
\begin{equation}
  W_{\mathrm{flux}}=\int_X G_3\wedge \Omega,
\end{equation}
where $\Omega$ is the Calabi--Yau $(3,0)$--form. This superpotential depends on the dilaton through the flux, and on the complex structure moduli through $\Omega$, but since there is no dependence on the K\"ahler moduli, one must look for other means to stabilize them. Here, non--perturbative effects yield such a means. There are two possible origins of such a non--perturbative superpotential: Euclidean D3--brane instantons wrapping 4--cycles $D \subset X$ and gaugino condensation in the world-volume of D7--branes wrapping again 4--cycles $D\subset X$. Both give rise to terms in the superpotential of the form
\begin{equation}
  W_{\mathrm{np}}\sim g_i\,e^{-a\,V_i},
\end{equation}
where $V_i$ is the volume of the divisor $D_i$ wrapped by the brane and therefore depends on the K\"ahler moduli. The prefactor $g_i$ comes from a one--loop determinant and in general depends on the complex structure moduli. Unfortunately, it is very hard to calculate explicitly, however recently a nice method has been presented in~\cite{Berglund:2005dm}. Otherwise, the best one can do in general is deciding whether it is zero or not. In the absence of fluxes Witten gave a criterion for this in the framework of F--theory~\cite{Witten:1996bn}: The Calabi--Yau threefold $X$ is lifted to a Calabi--Yau fourfold $Y$ and the divisor $D \subset X$ is lifted to divisor $\widetilde{D} \subset Y$. Then the condition is that the holomorphic Euler characteristic of $\widetilde{D}$ has to equal one:
\begin{equation}
  \label{eq:chi}
  \chi({\cal O}_{\widetilde{D}})=\sum_{i=0}^3 h^{0,i}(\widetilde{D}) = 1.
\end{equation}
Under favorable circumstances, this condition can be reduced to a condition on the holomorphic Euler characteristic of the divisor $D$. The Hodge numbers $h^{0,1}(D)$ and $h^{0,2}(D)$ are also relevant to decide whether there are contributions from D7-branes~\cite{Gorlich:2004qm}.


\section{Moduli stabilization for resolved toroidal orbifold models}
\label{sec:Moduli}

We want to know whether the KKLT scenario can actually be realized, and in a first step, one is trying to understand the problem of moduli stabilization. One way to gain a better understanding is to construct specific examples which fit all requirements. While the complex structure moduli and the dilaton can be generically fixed by fluxes,  the first impression was that divisors which generate contributions to the non-perturbative superpotential are not generic~\cite{Witten:1996bn}, \cite{Robbins:2004hx}. 

In the search for models with all moduli fixed by the effects described above, we follow the example of~\cite{Denef:2005mm} and consider orientifolds of resolved toroidal orbifolds. The motivation to look at this specific class of models is that toroidal orbifolds have a simple structure which allows for many quantities to be computed explicitly. Furthermore, they are also phenomenologically attractive because of their standard model-like features such as non--abelian gauge groups, chiral fermions and family repetition.

Orbifolds come with a number of untwisted K\"ahler moduli, which are inherited directly from the underlying torus, plus a number of twisted K\"ahler moduli which parametrize the sizes of the exceptional divisors which result from the resolution of the orbifold singularities. If one takes the aim of fixing all moduli seriously, one has to take care of the twisted K\"ahler moduli as well. Most likely, they will not be stabilized at zero size, so one has to look at the smooth Calabi--Yau manifold resulting from all orbifold singularities being resolved. The resolution process is most conveniently described in the language of toric geometry.  Divisors originating from blow--ups are particularly promising candidates for yielding a non--perturbative superpotential since they are in general rigid, i.e. have no moduli.

Ultimately, we would like to check all toroidal orbifolds which allow tadpole cancellation in type IIB for their suitability for the KKLT proposal. The example of $T^6/ \mZ_2 \times \mZ_2$ which was studied in great detail in~\cite{Denef:2005mm} turned out to allow stabilization of all moduli. In the class of toroidal orbifolds, this model holds a special place because of its simplicity (the orbifold group contains only  three non--trivial elements, which give rise to only one sort of local patch) and symmetry (the configuration of fixed sets as well as of the O--planes is the same for every coordinate direction) and its F--theory lift is known. This is radically different for most other toroidal orbifolds as we will explain in the following sections. For this reason we would like to ascertain how common the stabilization of all moduli in the class of toroidal orbifolds is. Furthermore, we believe that the construction of blowing up the various different patches and gluing them together to form a smooth Calabi--Yau manifold is worthwhile to be understood in its own right.

In the following, we will sketch the recipe of~\cite{Denef:2005mm} and go on to give some of the steps explicitly for our example.
\begin{itemize}
\item We begin with some toroidal orbifold. The regions close to the orbifold singularities can be described with toric geometry. In these local patches, we can resolve the singularities via a blow--up.
\item In the next step, those local patches have to be glued together to form a smooth, compact Calabi--Yau threefold $X$.
\item The topologies of the divisors of $X$ have to be analyzed. This will be important in order to decide which of them can generate a non--perturbative contribution to the superpotential.
\item A consistent orientifold projection has to be found and performed. This yields the O--planes and changes the geometry. The tadpoles can be cancelled by adding D--branes.
\item The K\"ahler potential has to be determined. For the K\"ahler part, we have to compute the intersection ring of $X$. This yields a non--trivial consistency check on the the previous steps.  
\item The 3--form flux is turned on to fix the complex structure moduli and the dilaton. The effect of the background flux on the fermionic zero modes on the D--branes and therefore the generation of the non--perturbative terms in the superpotential has to be taken into account.
\item In the last step, the scalar potential obtained from the superpotential and the K\"ahler potential has to be minimized.
\end{itemize}


\subsection{Toroidal Orbifolds}
\label{sec:orbifolds}

Probably the simplest way to construct a Calabi--Yau compactification is to take a toroidal orbifold and resolve its singularities. Such orbifolds have been analyzed in detail some time ago. This analysis resulted in a partial classification~\cite{Font:1988mk}, \cite{Erler:1992ki} which is our starting point. These orbifolds are of the form $T^6/G$ where $G$ is either $\mZ_n$ or $\mZ_n \times \mZ_m$, where $G$ acts by multiplying the complex coordinates by phases. The $G$ action has fixed sets at which the orbifold is singular. A point $f^{(n)}$ is fixed under $\theta^n\in G,\ \ n=0,...,|G|-1,$ if it fulfills 
\begin{equation}
  \label{eq:fix}
  \theta^n\,f^{(n)}=f^{(n)}+a,\quad a\in \Lambda,
\end{equation}
where $a$ is a vector of the torus lattice $\Lambda$. These singularities have to be resolved in order to obtain a smooth Calabi--Yau threefold and we review this in the next subsection.

As a specific example, we will discuss here $T^6/{\mZ}_{6-II}$ compactified on the root lattice of $SU(2)^2\times SU(3)\times G_2$.
In complex coordinates, the twist acts as
\begin{equation}
  \label{eq:action}
  \theta:\ (z^1, z^2, z^3) \to (\varepsilon\, z^1, \varepsilon^2\, z^2, \varepsilon^3\, z^3), \quad \varepsilon=e^{2\pi i/6}.
\end{equation}

\begin{figure}[h!]
\begin{center}
\includegraphics[width=80mm]{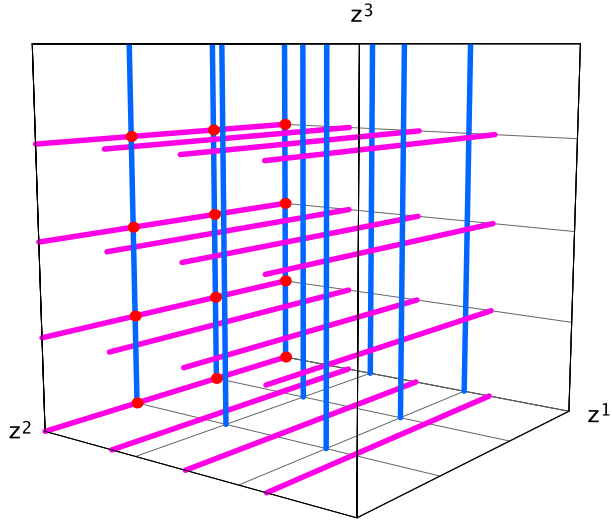}
\caption{Schematic picture of the fixed set configuration of $\mZ_{6-II}$ on $SU(2)^2\times SU(3)\times G_2$}
\label{fig:fix}
\end{center}
\end{figure}

\begin{table}
\label{tab:table}
\begin{center}
\begin{tabular}{@{}llll@{}}
\hline
Group element & Order & Fixed set & Conj. classes \\
\hline
$\theta$& 6 &12 fixed points & 12 \\
$\theta^2$ & 3 &9 fixed lines &  6 \\
$\theta^3$ & 2 &16 fixed lines & 8 \\
\hline
\end{tabular}
\end{center}
\caption{Fixed point set for $\mZ_{6-II}$ on $SU(2)^2\times SU(3)\times G_2$}
\end{table}
The fixed point sets are listed in Table~\ref{tab:table} and a schematic picture is displayed in Figure~\ref{fig:fix}. Each complex coordinate is shown as a coordinate axis and the opposite faces of the resulting cube are identified. The covering space is shown, not the quotient. 

Here we see a first important difference from the example in~\cite{Denef:2005mm}. The configuration of the fixed sets is different in the different coordinate directions and there are several different types of singularities. This is the typical situation for all the orbifold models we consider. Furthermore, it is not yet known how the F--theory lift can be explicitly constructed. From the fixed point configurations one finds the Hodge numbers to be $h^{1,1}=3+32=35$ and $h^{2,1}=1+10=11$. From these one can see that this manifold possibly allows for a Borcea--Voisin realization~\cite{Voisin:1993ab},~\cite{Borcea:1997tq}, from which the F-theory fourfold could be constructed. This is, however, not the case in general. If no F--theory lift is known, Witten's criterion~(\ref{eq:chi}) cannot be applied. A possible way out is the method presented in~\cite{Bergshoeff:2005yp}.


\subsection{Resolution of Singularities}
\label{sec:local}

A systematic way to resolve abelian orbifold singularities is toric geometry~\cite{Fulton:1993ab}. In the following paragraphs, we summarize some of the basic facts about toric geometry. An $n$--dimensional toric variety takes the form
\begin{equation}
  X_\Sigma=(\mC^d\setminus F_\Sigma)/(\mC^*)^r,
\end{equation}
where $n=d-r$, and $(\mC^*)^r$ acts by coordinatewise multiplication. The set $F_\Sigma$ is a subset that remains fixed under a continuous subgroup of $(\mC^*)^r$ and must be subtracted for the variety to be well defined. The action of $(\mC^*)^r$ is encoded in a lattice $N$ which is isomorphic to $\mZ^d$ and by its fan $\Sigma$, a collection of strongly convex rational cones in $N\otimes_\mZ \mR$ with the property that each face of a cone in $\Sigma$ is also a cone in $\Sigma$ and the intersection of two cones in $\Sigma$ is a face of each. The $k$--dimensional cones in $\Sigma$ are in one--to--one correspondence with the codimension $k$--submanifolds of $X_\Sigma$. In particular, the one--dimensional cones correspond to the divisors in $X_\Sigma$. The fan $\Sigma$ can be encoded by the generators of its edges or one--dimensional cones, i.e. by vectors $v_i\in N$. To each $v_i$ we associate a homogeneous coordinate $U_i$ of $X_\Sigma$. The $(\mC^*)^r$ action is encoded on the $v_i$ in $r$ linear relations
\begin{equation}
  \label{eq:linrels}
  \sum_{i=1}^d l^{(a)}_i v_i = 0, \qquad a=1,\dots,r, \quad l^{(a)}_i \in \mZ.
\end{equation}
We are only interested in Calabi--Yau orbifolds $\mC^m/G$ of dimensions $m=2,\,3$, so we require $X_\Sigma$ to have trivial canonical class. This translates to demanding that all but one of the $v_i$ lie in the same affine hyperplane one unit away from the origin $v_0$. This means that the last component of all the $v_i$ (except $v_0$) equals one. This allows us to draw toric diagrams in two (one) dimensions instead of $m=3$ ($m=2$). The fan $\Sigma$ associated to the former is obtained as follows: We have a single three--dimensional cone in $\Sigma$, generated by $v_1,\,v_2,\,v_3$. A generator $\theta$ of $G$ of order $n$ acts on the coordinates of $\mC^3$ by
\begin{equation}
  \theta:\ (z^1,\, z^2,\, z^3) \to (\varepsilon\, z^1, \varepsilon^{n_1}\, z^2, \varepsilon^{n_2}\, z^3),\quad \varepsilon=e^{2 \pi i/n},
\end{equation}
Then the local coordinates of $X_\Sigma$ are $U^k=(z^1)^{(v_1)_k}(z^2)^{(v_2)_k}(z^3)^{(v_3)_k}$. To find the coordinates in $N$ of the generators $v_i$ of the fan, we require the $U^k$ to be invariant under the action of $\theta$. This results in finding two linearly independent solutions of the equation
\begin{equation}
  \label{eq:vi}
  (v_1)_k+n_1\,(v_2)_k+n_2\,(v_3)_k= 0\ \mod\, n.
\end{equation}
The divisors corresponding to $v_i$ will be denoted by $D_i$.

$X_\Sigma$ is smooth if all the top--dimensional cones in $\Sigma$ have volume one. Here, there is only one such cone whose volume is $|G|$, hence $X_\Sigma$ is singular. There is a standard procedure for resolving singularities of toric varieties. It consists of adding all lattice points in $N$ which lie in the polyhedron in the affine hyperplane at distance one which is spanned by the generators $v_i$. For the fan $\Sigma$ this means that the corresponding one--dimensional generators $w_i$ are added to it and that it has to be subdivided accordingly. We denote the refined fan by $\widetilde\Sigma$. In~\cite{Aspinwall:1994ev} it is shown that these new generators can be related (in the case $m=3$) to the generators of $G$ as follows:   
\begin{equation}
  \label{eq:crit}
  w_i=g^{(i)}_1\,v_1+g^{(i)}_2\,v_2+g^{(i)}_3\,v_3, \qquad \qquad\sum_{k=1}^3 g^{(i)}_k=1,\quad  0\leq g^{(i)}_k<1.
\end{equation}
where $g^{(i)}=(g^{(i)}_1,\,g^{(i)}_2,\,g^{(i)}_3)\in (\mZ_n)^3$ represents the corresponding generator $\theta_i$. The corresponding exceptional divisors are denoted $E_i$.
The subdivision of the fan $\Sigma$ into $\widetilde\Sigma$ corresponds to a triangulation of the toric diagram. In general, there are several triangulations, and therefore several possible resolutions. They are all related via birational transformations. 

The case $m=2$ is even simpler. The singularity $\mC^2/\mZ_n$ is called a rational double point of type $A_{n-1}$ and its resolution is called a Hirzebruch--Jung sphere tree consisting of $n-1$ exceptional divisors intersecting themselves according to the Dynkin diagram of $A_{n-1}$.

In our example, we see from Table~\ref{fig:fix} that we have lines of $\mC^2/\mZ_2$ and $\mC^2/\mZ_3$ singularities, as well as singular points of the form $\mC^3/\mZ_6$. For the latter we take the action in~(\ref{eq:action}). Then~(\ref{eq:vi}) leads to the following three generators of the fan $\Sigma$:
\begin{equation}
{v_1=(-2,-1,1),\ v_2=(1,-1,1),\ v_3=(0,1,1).}
\end{equation}
The points inside the triangle $\langle v_1, v_2, v_3 \rangle$ lead to four additional generators of $\widetilde\Sigma$:
\begin{equation}
w_1=(0,0,1),\quad w_2=(0,-1,1),\quad w_3=(-1,0,1),\quad w_4=(-1,-1,1).
\end{equation}
By~(\ref{eq:crit}) they correspond to the generators $\theta^i$, $i=1,\dots,4$, of $\mZ_{6-II}$. In this example, there are five triangulations. 
\begin{figure}[h!]
\includegraphics[width=\linewidth]{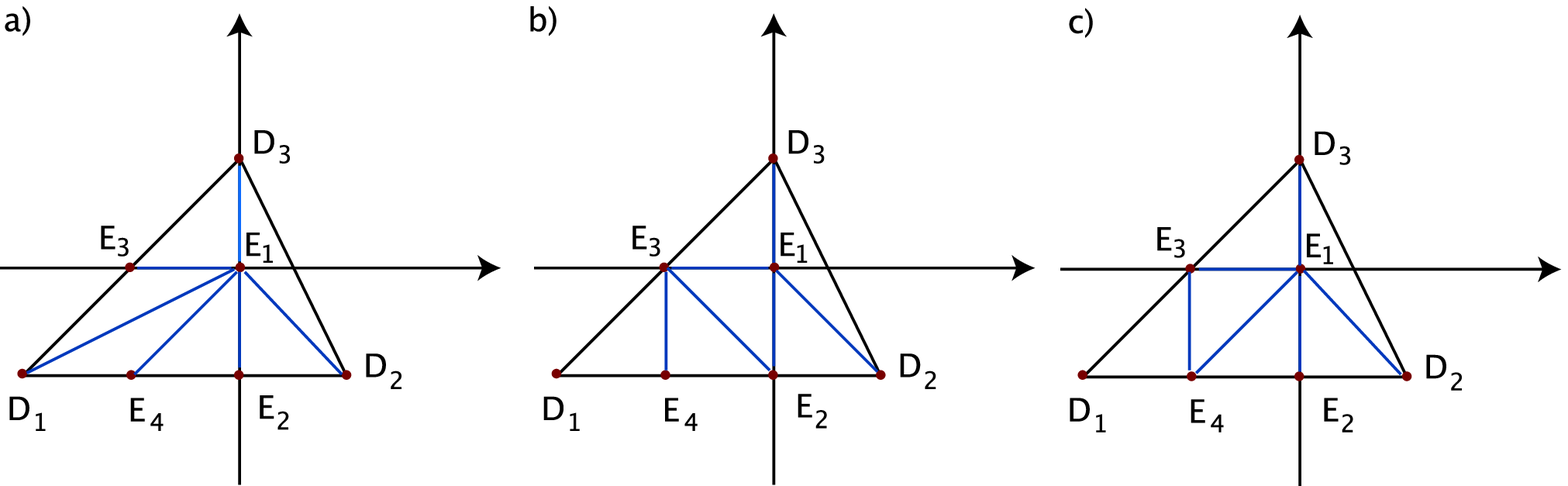}
\caption{Toric diagrams of the resolutions of $\mC^3/\mZ_{6-II}$}
\label{fig:diagram}
\end{figure}
Figure~\ref{fig:diagram} shows three of the corresponding toric diagrams. The remaining two not shown here are obtained by (i) taking the second case and blowing the curve $E_1\cdot E_3$ down and then blowing the curve $E_2\cdot D_2$ up (such a process is called a flop) and (ii) taking the third case and flopping the curve $E_1\cdot E_2$ to $D_3\cdot E_4$.


\subsection{The global resolution}
\label{sec:global}

Having analyzed the local structure of the singularities and their resolution we can return to the resolved global orbifold. The general procedure consists of gluing all the resolved local patches $X_{\widetilde{\Sigma}}$. This is achieved by relating their non--compact divisors $D_i$ to the divisors $R_i$, which are inherited from the covering torus $T^6$, via the linear relations of the toric varieties $X_{\widetilde{\Sigma}}$.

The inherited divisors and the exceptional divisors together form a basis for the divisor classes of the resolved orbifold. In addition, there are further natural divisors which will be expressed in terms of linear combinations of these basis elements. These additional divisors are simply planes lying at fixed loci of the orbifold action: $D_{i\alpha} = \{ z^i = \zf{i}{\alpha} \}$ where $\alpha$ runs over the fixed loci in the $i$th direction. In terms of the local toric patches they correspond to the non-compact divisors $D_i$. The local linear relations become ``sliding'' divisors in the compact geometry, and are related to the inherited divisors as follows. Consider the divisors $\{ z^i = c \not = \zf{i}{\alpha} \}$. They ``slide'' in the sense that they can move away from the fixed point, and the way they can move is constrained by the linear relations in the local geometry. We need, however, to pay attention whether we use the local coordinates $\tilde z^i$ near the fixed point on the orbifold or the local coordinates $z^i$ on the cover. Locally, the map is $\tilde z^i = \left(z^i\right)^n$, where $n$ is the order of the group element that fixes the point. The divisor $R_i=\{ \tilde z^i = c^n \}$ on the orbifold lifts to a union of $n$ divisors $R_i = \bigcup_{k=1}^n \{ z^i = \varepsilon^k c\}$ with $\varepsilon^n =1$. Consider the local toric patch before blowing up. The fixed point lies at $c=\zf{i}{\alpha}$ and in the limit as $c$ approaches this point we find the relation between $R_i$ and $D_i$ to be $R_i \sim nD_i$. This expresses the fact that at the fixed point the polynomial defining $R_i$ on the cover has an $n$th order zero on $D_i$. In the local toric patch $R_i \sim 0$, hence $nD_i \sim 0$. After blowing up, the $R_i$ and $nD_i$ differ by the exceptional divisors $E_k$ introduced in the process of resolution. The difference is expressed precisely by the linear relation~(\ref{eq:linrels}) corresponding to the resolved toric variety $X_{\widetilde\Sigma}$ and takes the form $R_i \sim nD_i + \sum_k E_k$. From this we see that this relation is independent from the chosen resolution. Such a relation holds for every fixed point $\zf{i}{\alpha}$ which adds the label $\alpha$ to the relation:
\begin{equation}
  \label{eq:Ri}
  R_i \sim nD_{i\alpha} + \sum_k E_{k\alpha}.
\end{equation}
The precise form of the sum over the exceptional divisors depends on the singularities involved.

\begin{figure}
\includegraphics[width=50mm]{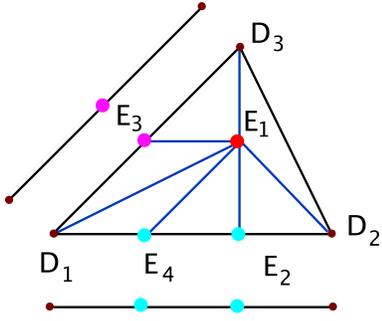}%
\caption{Toric diagrams of the resolution of the $\mC^3/\mZ_{6-II}$ fixed point in $T^6/\mZ_{6-II}$. We emphasize that the points in the interior of the boundary are identified with the Dynkin diagrams of $A_1$ and $A_2$ obtained from resolving the $\mC^2/\mZ_2$ and $\mC^2/\mZ_3$ fixed lines which intersect in this $\mC^3/\mZ_{6-II}$ fixed point.}
\label{fig:glue}
\end{figure}

For our example, we obtain the following: From Figure~\ref{fig:fix} we see that we have 12 local $\mC^3/\mZ_{6-II}$ patches which each sit at the intersection of a $\mC^2/\mZ_2$ and a $\mC^2/\mZ_3$ fixed line. They each have one exceptional divisor in the interior and three on the boundary of the toric diagram. From the discussion in Section~\ref{sec:local} we see that the 6 $\mC^2/\mZ_3$ fixed lines each contribute two exceptional divisors. At the $\mC^3/\mZ_{6-II}$ fixed point sitting on this line, they are identified with the two divisors sitting on the boundary $\langle D_1,D_2\rangle$. Then, there are 8 lines fixed under the $\mZ_2$--element, contributing each one exceptional divisor. In the $\mC^3/\mZ_{6-II}$ fixed point sitting on top of such a fixed line, the divisor is identified with the one on the boundary $\langle D_1,D_3 \rangle$ of the toric diagram of the resolution, as before. We have illustrated this in Figure~\ref{fig:glue}. The vertices corresponding to the exceptional divisors are colored to match Figure \ref{fig:fix}. This gives a total of $12\cdot 1+6\cdot 2+8\cdot 1=32$ exceptional divisors. In addition, we have three inherited divisors $R_i = \{\tilde z^i = c^6\},\ i=1,2,3$ which yields $h^{1,1}=35$. 

Again, from Figure~\ref{fig:fix} we see that the divisors located at fixed sets are three planes $D_{2,\beta}$ at the $\mZ_3$ fixed lines and four planes $D_{3,\gamma}$ at the $\mZ_2$ fixed lines. In the $z^1$ direction some of the fixed lines get identified by $\theta$, so that there remain three equivalence classes $D_{1,\alpha}$. As representatives we choose $z^1=0$ for $\alpha=1$, $z^1=1/2$ for $\alpha=2$, and $z^1=1/3$ for $\alpha=3$. So for $E_{3,\gamma\alpha}$, we can have $\alpha=1, 2$ and for $E_{2/4,\beta\alpha}$, we can have $\alpha=1,3$. Using~(\ref{eq:crit}) in~(\ref{eq:linrels}), the relations~(\ref{eq:Ri}) become
\begin{eqnarray}
  \label{eq:relations}
  R_1&=&6\,D_{{1,1}}+3\,\sum_{\gamma=1}^4 E_{{3,\gamma1}}+\sum_{\beta=1}^3\sum_{\gamma=1}^4E_{{1,\beta\gamma}}+\sum_{\beta=1}^3[\,2\,E_{{2,\beta1}}+4\,E_{{4,\beta1}}],\cr
  R_1&=&6\,D_{{1,2}}+3\,\sum_{\gamma=1}^4 E_{{3,\gamma2}},\cr
  R_1&=&6\,D_{{1,3}}+\,\sum_{\beta=1}^4\, [2\,E_{{2,\beta3}}+E_{{4,\beta3}}],\\
  R_2&=&3\,D_{{2,\beta}}+\sum_{\gamma=1}^4E_{{1,\beta\gamma}}+\sum_{\alpha=1,3}[2\,E_{{2,\beta\alpha}}+E_{{4,\beta\alpha}}],\quad \beta=1,2,3,\cr 
  R_3&=&2\,D_{{3,\gamma}}+\sum_{\beta=1}^3E_{{1,\beta\gamma}}+\sum_{\alpha=1,2}E_{{3,\gamma\alpha}},\quad \gamma=1,\dots,4.\nonumber
\end{eqnarray}
These relations encode the gluing of the local patches and will be crucial in determining the intersection ring of the orientifold.


\subsection{Topology  of the divisors}
\label{sec:topology}

At this point, we can already determine the topology of the divisors in order to decide whether they can contribute to the superpotential (after the lift to F--theory). This can be done straightforwardly for all the cases.

The topology of the exceptional divisors can be discussed in the local toric variety $X_{\widetilde\Sigma}$. For that purpose we recall the notion of the star of a cone $\sigma$, denoted ${\rm Star}(\sigma)$, which is the set of all cones $\tau$ in the fan $\widetilde\Sigma$ containing $\sigma$. In our situation, the topology of an exceptional divisor $E_i$ is then determined in terms of ${\rm Star}(\sigma_{w_i})$. This means that we simply remove from the fan $\widetilde\Sigma$ all cones, i.e. points and lines in the toric diagram, which do not contain $w_i$. It can be shown that the compact exceptional divisors (i.e. those in the interior of the toric diagram) can either be a $\mP^2$, a Hirzebruch surface $\mF_n$, a toric blow--up thereof, or a flop of any of these possibilities. For the exceptional divisors on the boundary of the diagram the geometry is effectively reduced by one dimension. The only compact toric manifold in one dimension is $\mP^1$, the extra non--compact direction is $\mC$, and there is again the additional possibility of toric blow--ups and flops.

After the gluing process described in Section~\ref{sec:global} the non--compact factor is identified with a fixed line $T^2/\mZ_n$. After the resolution, the topology becomes a $\mP^1$ blown up at some points. Hence, the topology of these divisors is a (possibly non--trivial) product of two blown--up $\mP^1$s. The same argument also applies to the locally non--compact divisors $D_i$. In the global orbifold they take the form $T^2/\mZ_n \times T^2/\mZ_m$. With the singularities removed, this is the same as a (possibly non--trivial) product of two $\mP^1$s with a certain number of points removed in each factor. Blowing up the singularities glues in either points or further $\mP^1$s depending on the number of exceptional curves ending in $D_i$ in the toric diagram. 

The upshot is that all the divisors $D_i$ and $E_k$ are rational surfaces which have in particular $h^{1,0} = h^{2,0} = 0$. The inherited divisors $R_i$ naturally come with the position modulus $c$, i.e. have $h^{1,0}\not=0$.

In our example, the only compact exceptional divisor is $E_1$. In the triangulation b), we recognize its star to be that of an $\mF_1$. In the other triangulations, $E_1$ is birationally equivalent to $\mF_1$, since the triangulations are all connected to each other through flops. As for the locally non--compact exceptional divisors, in triangulation a) they all have the topology of $\mP^1\times \mC$, while in the other triangulations, we have first to perform blow--ups and flops to reduce them to this simple form.

\subsection{Orientifold projection}
\label{sec:orientifold}

In order to obtain a consistent theory, an orientifold projection has to be performed. The easiest way would be to start with the corresponding orientifold at the orbifold point and then blowing up. However, in this process non--perturbative states from D--branes wrapping vanishing cycles may appear, making perturbation theory singular and thus preventing the orbifold phase and the smooth phase from being  continuously connected~\cite{Denef:2005mm} (for explicit examples in a slightly different context, see~\cite{Brunner:2004zd}). Therefore, we have to resolve the singularities first and impose the orientifold projection directly on the smooth geometry.  Since we are interested in the situation with O3-- and O7--planes, we choose the geometric part of the orientifold action $\Omega I_6$ on the torus $T^6$ to be $I_6:\ (z^1,z^2,z^3)\to (-z^1,-z^2,-z^3)$. We therefore have to study the fixed sets under the combination of $I_6$ and the elements of the orbifold group $G$. The condition~(\ref{eq:fix}) is replaced by
\begin{equation}
  I_6\,\theta^i\,z=z+a,\quad a\in \Lambda,\ i=0,...,n-1.
\end{equation}

In contrast to the simple example in~\cite{Denef:2005mm}, these fixed point sets in general do not coincide with the fixed point set of the orbifold group. In particular, there are orbifold fixed points which are mapped into each other by the orientifold action. Consequently, the orientifold action is locally trivial. Furthermore, in the split of $H^{1,1}(X) = H^{1,1}_+(X) \oplus H^{1,1}_-(X)$ into an invariant and an anti--invariant part, we now have $h^{1,1}_- \not = 0$. The corresponding moduli $G^a$ are not K\"ahler moduli anymore but have the form~\cite{Grimm:2005fa}
\begin{equation}
  \label{eq:G}
  G^a = C_2^a - S\, B_2^a.
\end{equation}
Currently, there is no superpotential known which could fix these moduli~\footnote{We thank Thomas Grimm for pointing this out to us.}. 

In addition, we have to decide whether and how we impose an orientifold action on the local coordinates $y_k$ of the exceptional divisors $E_k$. Presently, no criteria are known how to do this except for consistency at the very end of the computations. This leaves a lot of freedom. It turns out that indeed a non--trivial action on the $y_k$ is needed. In order to emphasize the importance of this point, we mention two places where this choice becomes important. When a divisor is fixed by the orientifold action, a factor of 1/2 appears in front this divisor in~(\ref{eq:Ri}) resp.~(\ref{eq:relations}). Moreover, although the divisors of all models in question have similar properties, namely $h^{1,0}=h^{2,0}=0$, it depends strongly on the choice of orientifold involution whether such a divisor will contribute to the non--perturbative superpotential or not.


\section{Outlook}
\label{sec:outlook}

Of the seven steps outlined in the beginning of section \ref{sec:Moduli}, we have addressed only the first four in this article. The remaining three, such as the calculation of the K\"ahler potential, turning on 3--form flux and the minimization of the scalar potential will be addressed in~\cite{toappear}, also for the other examples discussed in~\cite{Lust:2005dy}. 

We can make a preliminary comment on the second but last step of turning on the 3--form fluxes. Since most of the divisors in our model have vanishing $h^{1,0}$ and $h^{2,0}$, the background flux will have no effect on the zero modes. The reason is that the flux can lift zero modes. Since $h^{1,0}=h^{2,0}=0$, it can only have an effect on $h^{0,0}$, which is 1 for connected surfaces. But as long as we turn on supersymmetric fluxes, $h^{0,0}$ is left untouched. 

As a further step a complete formulation of these models in F--theory would be useful in order to apply the methods of~\cite{Berglund:2005dm} to explicitly compute the non--perturbative contributions to superpotential as well as to honestly take into account quantum corrections by going away from the constant dilaton limit. The simultaneous simplicity and non--triviality of these models make them promising candidates for such investigations.
\\

\vspace{2cm}
\noindent{\bf Acknowledgments}

We would like to thank Dieter L\"ust, Stephan Stieberger and Waldemar Schulgin for collaboration on the project~\cite{toappear}, for which this article is a preview. Furthermore, we would like to thank Frederik Denef, Bogdan Florea and Thomas Grimm for helpful discussions and correspondence. This article will appear in the proceedings of the RTN meeting "Constituents, Fundamental Forces and Symmetries of the Universe", in Corfu, Sep. 20-26, 2005. S.R. would like to thank the organizers of this workshop for providing such a pleasant atmosphere. S.R. would like to thank the University of Munich for hospitality. E.S. is supported by the Marie Curie Grant MERG--CT--2004--006374.

\end{document}